
\magnification=\magstep1
\baselineskip=18pt
\overfullrule=0pt
\nopagenumbers
\footline={\ifnum\pageno>1\hfil\folio\hfil\else\hfil\fi}
\font\twelvebf=cmbx12

\rightline{CU-TP-601}
\rightline{June 1993}
\vskip .8in
\centerline {\twelvebf Hard Thermal Loops, Gauged WZNW-Action and the
Energy of }
\centerline{\twelvebf Hot Quark-Gluon Plasma}
\vskip .8in
\centerline{\it V.P.Nair $^*$}
\smallskip
\centerline{Physics Department, Columbia University}
\centerline{New York, NY 10027}
\vskip 1in
\centerline{\bf Abstract}
\smallskip
\noindent
The generating functional for hard thermal loops in Quantum
Chromodynamics is rewritten in terms of
a gauged Wess-Zumino-Novikov-Witten action
by introducing an auxiliary field. This shows in a simple way that the
contribution of hard thermal loops to the energy of the quark-gluon
plasma is positive.
\vskip 1in
\noindent This research was supported in part by the U.S. Department
of Energy.
\footnote {}{$^*$ Address after September 1, 1993: Physics Department,
City College of the CUNY, Convent Avenue at 138 Street, New York, NY
10031.}
\vfill\eject
\def\h{{\textstyle {1\over 2}}}
\def\14{{\textstyle {1\over 4}}}
\def\cS{{\cal S}}
\def\cM{{\cal M}}
\def\cA{{\cal A}}
\def\cH{{\cal H}}
\def\tr{{\rm tr}}
\def\x^T{{{\bf x}^T}}

The effective action or generating functional $\Gamma$ for hard thermal
loops in Quantum Chromodynamics (QCD) has been the subject of many
recent investigations [1-4]. We have shown that $\Gamma$ is essentially given
by the eikonal for a Chern-Simons (CS) gauge theory [3]. $\Gamma$ is a
nonlocal functional of the gauge potential $A_\mu$ and leads to new
effective propagators and vertices. These effective propagators and
vertices help to carry out the reorganization (and partial summation) of
thermal perturbation theory necessary to ensure that all contributions
to a given order in the coupling constant are consistently included.
Further, many of the properties of the quark-gluon plasma at high
temperature can be understood in terms of the action
$$
{\cal S}= \int ~\14  F^2 ~+~\Gamma [A] \eqno(1)
$$
where we add $\Gamma$ to the usual Yang-Mills action. The action (1)
gives a gauge invariant description of Debye screening and plasma waves.
Also, with a proper $i\epsilon$-prescription, the equations of motion
following from (1) (which are also nonlocal) can describe Landau
damping [4]. Clearly further investigation of $\Gamma$ is interesting both
for the programme of resummation of perturbation theory and for
clarifying properties of the quark-gluon plasma at high temperature.

In this paper, we introduce an auxiliary field and rewrite $\Gamma$ in
terms of a gauged Wess-Zumino-Novikov-Witten (WZNW) theory. This field is
actually defined on a six-dimensional space. The elimination of
the auxiliary field via its equations of motion leads back to (1) with
$\Gamma$ expressed in terms of the potentials. The introduction of the
auxiliary field has the advantage that the equations of
motion are {\it local}. This facilitates further analysis since the
quark-gluon plasma (at high temperature) can now be described by a gauge
theory coupled, in a local and more or less standard way, to an
auxiliary field. We also obtain the energy functional and show that it
is positive for all physical field configurations, a question which has
been of some concern recently [5]. This rewriting of the theory may also
help in setting up the resummed perturbation theory; new auxiliary
field propagators and vertices can be used to incorporate hard thermal
loop effects.

We begin by recalling some of the essential results of ref.[3]. $Q,~Q'$
will denote the null vectors
$Q_\mu =(1,~{\vec Q}),~~Q'_\mu =(1,-{\vec Q})
$
with $\vec Q\cdot \vec Q =1$. We introduce the light-cone coordinates
$u=\h (Q'\cdot x),~v=\h (Q\cdot x)$; the Euclidean version of $Q'\cdot
x$ and $Q\cdot x$ will be denoted by $z, \bar z$ respectively. The
corresponding components of the potential are $ A_+ = \h A\cdot Q,~A_-
=\h A\cdot Q',$ with $A_+ \rightarrow A_z,~A_-\rightarrow A_{\bar z}$
in the Euclidean case. Here $A_\mu= -it^a A^a_\mu$; $t^a$ are hermitian
matrices
representing the generators of the Lie algebra of the gauge group,
which we take to be $SU(N)$.

We shall, for simplicity of presentation, introduce the auxiliary field
in its Euclidean version first. The effective action $\Gamma$ can be
written as
$$
\Gamma = ~-{k\over \pi} \int d\Omega ~d^2 \x^T~
\left[ \int d^2z~\tr(A_z A_{\bar z})~+~ i\pi I(A_z)
{}~+~i\pi I(A_{\bar z})\right] \eqno(2)
$$
where
$k= (N+ \h N_F ){T^2\over 6}$;
$N_F$ is the number of quark flavors and $T$ is the temperature.
The integration $\int d\Omega$ in (2) is over the angular directions of
$\vec Q$. $\x^T$ denotes the two coordinates transverse to $\vec Q$, i.e.
$\vec Q\cdot \x^T=0$.
$I(A_z)$ in (2) is the eikonal for a CS theory defined by
$$
I= ~i\sum_2^\infty ~{(-1)^n\over n} \int {d^2z_1\over \pi}\cdots
{d^2z_n\over \pi}~{{\tr (A_z (x_1)\cdots A_z(x_n))}\over {\bar
z}_{12}\cdots {\bar z}_{n-1n}~{\bar z}_{n1}}\eqno(3)
$$
(${\bar z}_{ij}= {\bar z}_i -{\bar z}_j$.) All $A$'s have the same
argument for the transverse coordinates $\x^T$. $I(A_z)$ has the
property that
$$
\delta I(A_z)= {i\over \pi}~\int d^2z~ \tr (a_{\bar z}~\delta A_z)\eqno(4a)
$$
where $a_{\bar z}$ obeys the equation
$$
\partial_z a_{\bar z}-\partial_{\bar z} A_z +[A_z, a_{\bar
z}]=0\eqno(4b)
$$
$I(A_{\bar z})$ in (2) is obtained from $I(A_z)$ by
$Q\leftrightarrow Q'$.
Equation (4b) has the following solution for $a_{\bar z}$.
$$
a_{\bar z}= \sum_1^\infty ~(-1)^{n-1}\int {d^2z_1\over \pi}\cdots
{d^2z_n\over \pi}~{{A_z(x_1)\cdots A_z(x_n)}\over {({\bar z}-{\bar
z}_1){\bar z}_{12}\cdots ({\bar z}_n-{\bar z})}}\eqno(5)
$$

$I(A_z)$ can also be related to the chiral Dirac determinant in two
dimensions and to the WZNW-action [6]. It is possible to introduce an
$SL(N,{\bf
C})$-matrix $U$ such that $A_z= -\partial_z U~U^{-1}$ and $A_{\bar
z}\equiv -(A_z)^\dagger = U^{\dagger ~-1} \partial_{\bar z} U^\dagger.$
The eikonal $I(A_z)$ can then be written as $I= i\log \det
(\partial_z +A_z)~= -i S(U)$. $S(U)$ is the WZNW-action [7]
$$
S(U)= {1\over 2\pi}\int d^2z ~\tr (\partial_z U\partial_{\bar
z}U^{-1}) -{i\over 12\pi} \int d^3x ~\epsilon^{\mu\nu\alpha}
\tr (U^{-1}\partial_\mu U ~U^{-1}\partial_\nu U~U^{-1}\partial_\alpha
U)\eqno(6)
$$
$S$ obeys the Polyakov-Wiegmann (PW) composition rule [6]
$$
S(hg)= S(h)~+~S(g)~-~ {1\over \pi}\int d^2z~
\tr (h^{-1}\partial_{\bar z} h~ \partial_zg g^{-1})\eqno(7)
$$
In terms of $U$, we may in fact write $\Gamma$ as
$$
\Gamma = -k \int
d\Omega ~d^2 \x^T~ S(U^\dagger U)\eqno(8)
$$

Using $\Gamma$ as given by (2) in the action (1), we can write down the
equations of motion for the gauge field. (Equation (4a) or the
PW-property (7) can be used for the variation of the action.) We find
the equations of motion
$$
(D_\mu F^{\mu\nu})^a ~+~ {k\over 2\pi}\int d\Omega~ \tr \left\{(-it^a)\left[
(a_{\bar z} -A_{\bar z})Q^\nu ~+~ (a_z-A_z)Q'^\nu \right]\right\} =0\eqno(9)
$$
(This equation was given in ref.[4].) Notice that we
may regard $a_{\bar z}, ~a_z$ as
new fields which are determined by equation (4b) and its conjugate.
The solution (5) and its conjugate give the current in (9).

We now introduce an auxiliary field $M(x,{\vec Q})$ which is a complex
$N\times N$ matrix of unit determinant. It is defined on $\cM\times S^2$
where $\cM$ is spacetime and $S^2$ is the two-sphere
corresponding to the orientations of $\vec Q$. In other words
$M(x,{\vec Q}):\cM\times S^2 ~\rightarrow SL(N,{\bf C})$.
We also define $H=M^\dagger M$.

The action for the fields $A^a_\mu (x),~M(x,{\vec Q})$ is taken to be
$$
\cS= \int ~\14 F^2 ~+~ k \int d\Omega ~d^2 \x^T~ {\tilde S(M,A)}\eqno(10)
$$
${\tilde S(M,A)}$ is the gauged WZNW-action given by [8]
$$
{\tilde S(M,A)}= S(H) + {1\over \pi}\int d^2z~ \tr ( H^{-1}\partial_{\bar
z}H~A_z- A_{\bar z} \partial_z H~H^{-1} ~+~ A_zH^{-1} A_{\bar
z}H-A_zA_{\bar z})\eqno(11)
$$
${\tilde S(M,A)}$ is invariant under the gauge transformation
$$
M\rightarrow M'= M~h^{-1},~~~~~~~~~~H\rightarrow H'=h~H~h^{-1}
$$
$$
A_\mu\rightarrow A'_\mu = h~A_\mu~ h^{-1}~-\partial_\mu h~h^{-1}\eqno(12)
$$
where $h\in SU(N)$ is a function of $x$ (not ${\vec Q}$). ${\tilde S(M,A)}$
can
be written in terms of $U$ defined by $A_z=-\partial_zU~U^{-1}$, by using
the PW-property (7), as
$$
{\tilde S(M,A)}= S(U^\dagger HU)~-~ S(U^\dagger U)\eqno(13)
$$
This makes the gauge invariance property manifest.

We shall now show that the elimination of the field $M(x,{\vec Q})$
leads back to the equation of motion (9) and the expression (8) for
$\Gamma$. The equations of motion for the effective action (10,11) can
be easily obtained using (7). We find
$$
\partial_z \cA_{\bar z}-\partial_{\bar z}A_z ~+~[A_z, ~\cA_{\bar z}]=0
\eqno(14a)
$$
$$
\partial_{\bar z}\cA_z -\partial_z A_{\bar z}~+~[ A_{\bar z},~\cA_z ]=0
\eqno(14b)
$$
$$
(D_\mu F^{\mu\nu})^a ~+~ {k\over 2\pi}\int d\Omega ~\tr \left\{
(-it^a)\left[ (\cA_{\bar z}-A_{\bar z})Q^\nu~+~ (\cA_z-A_z)Q'^\nu\right]
\right\}=0
\eqno(15)
$$
where
$$
\cA_z= H~A_z ~H^{-1}~-~ \partial_zH~H^{-1}
$$
$$
\cA_{\bar z}= H^{-1}~A_{\bar z}~H~+~ H^{-1}\partial_{\bar z}H\eqno(16)
$$
Equations (14a,14b) are in fact the same. They can be solved by
$\cA_{\bar z}\equiv  H^{-1}A_{\bar z}H+H^{-1}\partial_{\bar z} H= a_{\bar z}$,
the power series solution of (5). (This must be regarded as a solution
for $M(x,{\vec Q})$ or $H$. The explicit form for $H$ will be fairly
complicated; we do not need it.) Using this solution, we see that (15)
does indeed reduce to (9), showing that the action (10) is an
acceptable rewriting of the theory of equations (1,2). In terms of the
matrix $U$, the solution for $H$ is $U^{-1}U^{\dagger ~-1}$ or
$ M(x,{\vec Q})= U^{\dagger ~-1}$. Using this in (13), we see that
$$
\cS=\int~\14 F^2 ~-~ k\int d\Omega ~d^2 \x^T~S(U^\dagger U)\eqno(17)
$$
recovering $\Gamma$ in the form (8).

The equation of motion for the gauge field, viz. equation (15), can
also be written as
$$
(D_\mu F^{\mu\nu})^a ~+~ {k\over 2\pi}\int d\Omega ~\tr \left\{ (-it^a)\left[
H^{-1}D_{\bar z}H~Q^\nu - D_z H~H^{-1} ~Q'^\nu\right]\right\} =0\eqno(18)
$$
where $D_\mu$ is the gauge covariant derivative, $D_\mu H= \partial_\mu
+[A_\mu ,H]$.

The action (11) involves the Wess-Zumino (WZ) term, the second term on the
right hand side of (6) with $U\rightarrow H$,
which requires an extension of $H$ into a higher
dimensional space ${\tilde \cM}$. However, as discussed in ref.[3],
the topology of hermitian matrices being trivial, there is no argument
for the quantization of the coefficient $k$. In particular, we can find
a parametrization for $H$, e.g. $H=e^\theta,~\theta$ being hermitian and
traceless, such that the WZ-term can be written as an integral over
spacetime rather than ${\tilde \cM}$. Put another way, $\theta$ can be
thought of as $\log H$. In general, there are ambiguities related to the
choice of the branch of the logarithm which can eventually lead to the
quantization requirement on $k$. In the present case, the hermiticity of
$\theta$ eliminates the branch ambiguities.

We now consider the auxiliary field and action in Minkowski space. In
the Euclidean case, the solution for $H$ is of the form
$U^{-1}U^{\dagger~-1}$ where $A_z=-\partial_zU~U^{-1},~A_{\bar
z}=-\partial_{\bar z}(U^{\dagger~-1})~U^{\dagger}$. With Minkowski
signature, we can write
$$
A_+= -\partial_+ V~V^{-1},~~~~~~~~~~~A_-=-\partial_- V'~V'^{-1}\eqno(19)
$$
Evidently, $V'(x,{\vec Q})= V(x, -{\vec Q})$. $V(x,{\vec Q})$ can be
taken to be unitary. The comparison with the Euclidean version suggests
the introduction of an auxiliary field $N(x,{\vec Q})$ with
$G=N^{\dagger}(x,-{\vec Q})N(x,{\vec Q})$ playing the role of $H$.
The correspondence is thus
$$
M\leftrightarrow N(x,{\vec
Q}),~~~~~~~M^{\dagger}\leftrightarrow N^{\dagger}(x,-{\vec
Q}),~~~~~~~H\leftrightarrow G \eqno(20)
$$
$M(x,{\vec Q})$ is not unitary but $N(x,{\vec Q})$ is an
$SU(N)$-matrix.
$G$ is also special unitary and
from its definition is seen to satisfy the condition
$G^{\dagger}(x,{\vec Q})=G(x,-{\vec
Q})$.

The action in Minowski space is taken as
$$
\cS= \int -\14 F^2 +k\int d\Omega~\left[ \int d^2 \x^T~S(G)+ {1\over \pi}
\int d^4x~
\tr(G^{-1}\partial_- G~A_+ ~-~ A_- \partial_+ G~G^{-1}\right.
$$
$$
{}~~~~~~~~~~~~~~ +A_+ G^{-1}A_-
G-A_+ A_-)\Biggr]\eqno(21)
$$
where $S(G)$ is given by (6) with $U\rightarrow G$ and $\partial_+,
{}~\partial_-$ replacing $\partial_z, ~\partial_{\bar z}$.
The equations of motion are now given by
$$
\partial_+ \cA_- -\partial_- A_+ ~+~[A_+, \cA_-]=0\eqno(22a)
$$
$$
(D_\mu F^{\mu\nu})^a ~-~J^{\nu a}=0\eqno(22b)
$$
$$
J^{\nu a}= -{k\over 2\pi} \int d\Omega~ \tr \left\{ (-it^a)\left[
(\cA_- - A_-)Q^\nu ~+~(\cA_+ -A_+)Q'^\nu\right] \right\}
$$
$$
{}~~~~= -{k\over 2\pi} \int d\Omega~ \tr \left\{ (-it^a)\left[
G^{-1}D_-G~Q^\nu ~-~ D_+G ~G^{-1}Q'^\nu\right] \right\}\eqno(23)
$$
Again, solving equation (22a) and substituting in (22b), we can recover
the Minkowski version of equation (9). In solving equation (22a), we
need to specify certain analyticity properties or an
$i\epsilon$-prescription to define the inverse of $\partial_+$. As we
have discussed in ref.[4], this is a retardation property. (The
expression for the current in (23) is also similar to what was given in
the ref.[4]; the difference is that here $G$ is an independent field to
begin with.)

As in the Euclidean case,
the WZ-term seems to require an extension of $G$ into
a higher dimensional space. $G$ is now unitary and one might worry that
this would lead to a quantization requirement on $k$ so that physical
results remain the same for different extensions of $G$ into the higher
dimensional space. Actually we do not expect such a condition. $G$ is
not any unitary matrix; it has the property $G^{\dagger}(x,{\vec
Q})= G(x,-{\vec Q})$. Thus $\log G$ must be odd under ${\vec
Q}\rightarrow -{\vec Q}$. This condition will eliminate some of the
branch ambiguities so that
the WZ-term can again be written as an integral over
spacetime. Alternatively, existence of the Euclidean continuation via
rules (20) show that $k$ is not quantized.
Notice also that the fields
$A^a_\mu (x),~N(x,{\vec Q})$ obey {\it local} equations of motion.

We now turn to the energy functional. One could start with the action
(21), introduce phase space variables, Poisson brackets, etc. However,
there is a simpler way to obtain the energy functional. We consider the
action for a finite interval of time, say from zero to $t$, and use the
fact that the Hamiltonian or energy can be defined as
$\cH= -{\partial \cS\over \partial t}$.
As an example consider the theory of a scalar field $\phi$ with an
action
$$
\cS= \int_0^t dt~d^3x~ \left[ \h \left({\partial \phi\over \partial t}
\right)^2 -\h
(\nabla \phi)^2 -V(\phi )\right] \eqno(24)
$$
If we divide the interval $[0,t]$ into $n$ segments, with $t_0=0,~t_n=t$
and write
$$
\cS= \int d^3x~\left[ \h \sum {{(\phi_{n-i}-\phi_{n-i-1})^2}\over
{(t_{n-i}-t_{n-i-1})}}~- \left( \h (\nabla \phi)^2 +V(\phi
)\right)_{n-i} (t_{n-i}-t_{n-i-1})\right] \eqno(25)
$$
we see immediately that
$$
\cH \equiv -{\partial \cS\over \partial {t_n}} = \int d^3x~\left[
\h \left({\partial \phi\over \partial t}\right)^2 +\h (\nabla \phi)^2
+V(\phi )\right] \eqno(26)
$$
as $t_{n-i}-t_{n-i-1}\rightarrow 0, ~n\rightarrow \infty.$ This is of
course as expected. For the present case, the WZ-term in (21) has only
one time-derivative; it does not depend on $(t_n-t_{n-1})$ and so does
not contribute to the energy. From the rest of the terms in $\cS$, we
see immediately that
$$
\cH= \int d^3x~\left\{ {{E^2+B^2}\over 2}~+~ {k\over 8\pi}\int d\Omega~ \tr
\left[ (D_0G ~D_0G^{-1})+ ({\vec Q}\cdot {\vec D}G~ {\vec Q}\cdot {\vec
D}G^{-1})\right] -~ A^a_0{\cal G}^a\right\} \eqno(27)
$$
where $F^a_{0i}=E^a_i,~F^a_{ij}= \epsilon_{ijk}B^a_k$ and
$$
{\cal G}^a= ({\vec D}\cdot {\vec E})^a +{k\over 8\pi}\int d\Omega~ \tr
\left[ (-it^a)(G^{-1} D_- G~-~ D_+G~G^{-1})\right] \eqno(28)
$$

${\cal G}^a=0$ is the Gauss law of the theory. It is the time-component
of the equation of motion (22b). From (27), we see that $\cH$ is
positive for all physical field configurations, viz. for those
which satisfy the Gauss law.

In the action and the equations of motion, it is only the field
combination $G$ which appears. One could take this to be the basic field
instead of $N(x,{\vec Q})$, but the condition $G^\dagger (x,{\vec Q})=
G(x, -{\vec Q})$ has to be imposed so as to ensure that there are no
ambiguities in the extensions of $G$ required for the WZ-term.
\vskip .3in
\noindent{\bf References}
\vskip .1in
\item {1.} E.Braaten and R.Pisarski, {\it Phys.Rev.D} {\bf 42}, 2156
(1990); {\it Nucl.Phys.} {\bf B337}, 569 (1990); {\it ibid.} {\bf B339},
310 (1990); {\it Phys.Rev. D} {\bf 45}, 1827 (1992).
\vskip .1in
\item {2.} J.Frenkel and J.C.Taylor, {\it Nucl.Phys.} {\bf B334}, 199
(1990); J.C.Taylor and S.M.H.Wong, {\it Nucl.Phys.} {\bf B346}, 115
(1990).
\vskip .1in
\item {3.} R.Efraty and V.P.Nair, {\it Phys.Rev.Lett.} {\bf 68}, 2891
(1992); Columbia Preprint CU-TP 579, November 1992 (to be published in {\it
Phys.Rev.D}).
\vskip .1in
\item {4.} R.Jackiw and V.P.Nair, Columbia-MIT Preprint CU-TP 594,
CTP \# 2205, April 1993 (submitted to {\it Phys.Rev.D}).
\vskip .1in
\item {5.} F.T.Brandt, J.Frenkel and J.C.Taylor, Cambridge University
Preprint DAMTP-93-02, February 1993.
\vskip .1in
\item {6.} A.M.Polyakov and P.B.Wiegmann, {\it Phys.Lett.} {\bf 141B}, 223
(1984); D.Gonzales and A.N.Redlich, {\it Ann.Phys.} {\bf 169}, 104
(1986); G.V.Dunne, R.Jackiw and C.A.Trugenberger, {\it Ann.Phys.}
{\bf 194}, 197 (1989).
\vskip .1in
\item {7.} S.P.Novikov, Usp.Mat.Nauk {\bf 37}, 3 (1982);
E.Witten, {\it Commun.Math.Phys.} {\bf 92}, 455 (1984).
\vskip .1in
\item {8.} R.I.Nepomechie, {\it Phys.Rev.D} {\bf 33}, 3670 (1986);
D.Karabali, Q-H.Park, H.J.Schnitzer and Z.Yang, {\it
Phys.Lett.} {\bf 216B}, 307 (1989); D.Karabali and H.J.Schnitzer, {\it
Nucl.Phys.} {\bf B329}, 649 (1990);
K.Gawedzki and A.Kupianen, {\it
Phys.Lett.} {\bf 215B}, 119 (1988); {\it Nucl.Phys.} {\bf B320}, 649
(1989).
\end